\documentclass[preprint,prc,aps,epsfig]{revtex4}
\usepackage{graphicx,subfigure}
\usepackage{epsfig}

\begin{document}

\title{Kadomtsev-Petviashvili equation in relativistic fluid dynamics}

\author{D.A. Foga\c{c}a\dag\,  F.S. Navarra\dag\ and L.G. Ferreira Filho\ddag\ }
\address{\dag\ Instituto de F\'{\i}sica, Universidade de S\~{a}o Paulo\\
 C.P. 66318,  05315-970 S\~{a}o Paulo, SP, Brazil}
\address{\ddag\ Faculdade de Tecnologia, Universidade do Estado do Rio de Janeiro \\
Via Dutra km 298, CEP 27523-000, Resende, RJ, Brazil}

\begin{abstract}
The Kadomtsev-Petviashvili (KP) nonlinear wave equation is the three dimensional generalization of the
Korteveg-de Vries (KdV) equation. We show how to obtain the cylindrical KP (cKP) and cartesian KP in
relativistic fluid dynamics.  The obtained KP equations describe the evolution of perturbations in the baryon
density in a strongly interacting quark gluon plasma (sQGP) at zero temperature.
We also show the analytical solitary wave solution of the KP equations in both cases.

\end{abstract}


\maketitle


\vspace{1cm}
\section{Introduction}

The Standard Model (SM) is the well established and extremely successfull theory of the elementary particles and their
interactions \cite{text}.
According to the SM, matter is constituted by quarks and leptons and their interactions are due to the exchange of gauge bosons.
The part of the SM which describes the strong interactions at the fundamental level is called Quantum Chromodynamics, or QCD \cite{muta}.
In QCD the quarks of six types, or flavors, (up, down, strange, charm, bottom and top) interact exchanging gluons. Quarks and gluons have a
special charge called color, responsible for the strong interaction. Quarks and gluons do not exist as individual particles. Due to the
property of color confinement, quarks and gluons form clusters called hadrons, which can be grouped in baryons and mesons. The former are
made of three quarks,  as the proton, and the latter are made of a quark and an antiquark, as the pion. The quarks carry a fraction of the fundamental electric charge and a fraction of the baryon number. The electric and color charges and the baryon number are conserved quantities in QCD.

Under extreme conditions of very large temperatures and/or very large densities, the normal hadronic matter undergoes a phase transition to a
deconfined phase, a new state of matter called the quark gluon plasma, or QGP.  Together with the deconfinement phase transition, a second
phase transition takes place: the chiral phase transition, during which chiral symmetry is restored and the light quarks (up and down)
become massless.  The hot QGP is produced in relativistic heavy ion collisions
in the Relativistic Heavy Ion Collider (RHIC) at the Brookhaven National Laboratory (BNL) and even more in the Large Hadron Collider (LHC)
at CERN.  The cold QGP  may  exist in the core of  compact stars.

The discovery of QGP revealed also that it behaves as an almost perfect fluid and its space-time evolution can be very well described
by relativistic hydrodynamics. After the discovery of this new fluid, more sophisticated measurements made possible to study the propagation
of perturbations in the form of waves in the QGP. We may, for example, study the effect of a fast quark traversing the hot QGP medium. As is
moves supersonically throughout the fluid, it generates waves of energy density (or baryon density in the non relativistic case). In some works
it was even claimed that these  waves may pile up and form Mach cones \cite{jorge}, which would affect the angular  distribution of the
produced  particles, fluid fragments which are experimentally observed.

The study of waves in the quark-gluon fluid has been mostly performed with the assumption that the amplitude of the perturbations is small enough
to justify the linearization of the Euler and continuity equations \cite{shuryak}. As explained in the appendix, the analysis of perturbations with
the linearized relativistic  hydrodynamics  leads to the
standard second order wave equations and their travelling wave solutions, such as acoustic waves in the  QGP. While  linearization is justified in
many cases, in others it should be  replaced by another technique to treat perturbations keeping the nonlinearities of the theory.  This is where
a physical theory, in this case hydrodynamics, may benefit from developments in applied mathematics. Indeed, since long ago there is a technique which
preserves nonlinearities in the derivation of the differential equations which govern the evolution of perturbations. This is the
reductive perturbation method (RPM)\cite{leblond}.

In previous works we have applied the RPM to hydrodynamics  and we have shown that the nonlinearities may lead,
as they do in other domains of  physics, to new and interesting phenomena. In the case of the cold QGP we have shown \cite{nos2011a}
that it is possible to derive a Korteweg -  de Vries (KdV) equation for the baryon density, which has analytic solitonic solutions.
Perturbations in fluids with  different equations of state (EOS) generate different nonlinear wave equations: the breaking wave equation,
KdV, Burgers...etc. Among these equations we find
the Kadomtsev-Petviashvili (KP) equation  \cite{KPorigin}, which  is a nonlinear wave equation in three spatial and one temporal coordinate.
It is the generalization of the Korteweg-de Vries (KdV) equation to higher dimensions.  The KP equation describes the evolution of long waves of
small amplitudes with weak dependence on the transverse coordinates. This equation has been found with the application of the reductive perturbation
method \cite{rpm}
to several different problems such as the propagation of solitons in multicomponent plasmas, dust acoustic waves in hot dust plasmas and dense electro-positron-ion plasma
\cite{dassen,wen,jukui,kp2004,maiwen,yue1,yunliang,Wangetal,mushtaq,yue2,guang,kp2010,moslempp17}.

The main goal of this work is to apply the RPM \cite{rpm,dassen,wen,jukui,kp2004,maiwen,yue1,yunliang,Wangetal,mushtaq,yue2,guang,kp2010}
to relativistic fluid dynamics \cite{wein,land} in cylindrical and cartesian coordinates to obtain the KP equation.  We find that the transverse
perturbations in relativistic fluid dynamics may generate three dimensional solitary waves.

In the present study of relativistic hydrodynamics we shall consider an equation of state derived from QCD \cite{nos2011}.
The obtained energy density and pressure contain derivative terms
and a wave equation with a dispersive term such as KdV or KP emerges from the formalism. In \cite{nos2011a}, we have performed a similar study in
one dimension and found a KdV equation.  The present work is an extension of \cite{nos2011a} to three dimensions.

Previous studies on one-dimensional nonlinear waves in cold and warm nuclear matter can
be found in \cite{fn1,fn2,fn3,fn4,nos2010,frsw,abu}.

This text is organized as follows.
In the next section we review the basic formulas of relativistic hydrodynamics. In section III we derive the KP equation in
detail. In section IV we solve the KP equation analytically and in Section V we present some conclusions.

\section{Relativistic Fluid Dynamics}

For a detailed study in relativistic hydrodynamics
we suggest the references \cite{wein,land}.

The relativistic version of the Euler equation \cite{wein,land,nos2010,nos2011a} is given by:
\begin{equation}
{\frac{\partial {\vec{v}}}{\partial t}}+(\vec{v} \cdot \vec{\nabla})\vec{v}=
-{\frac{1}{(\varepsilon + p)\gamma^{2}}}
\bigg({\vec{\nabla} p +\vec{v} {\frac{\partial p}{\partial t}}}\bigg)
\label{eul}
\end{equation}
and the relativistic version of the continuity equation for the baryon density is
\cite{wein}:
\begin{equation}
\partial_{\nu}{j_{B}}^{\nu}=0
\label{conucleon}
\end{equation}
Since ${j_{B}}^{\nu}=u^{\nu} \rho_{B}$ the above equation can be rewritten as \cite{nos2010,nos2011a}:
\begin{equation}
{\frac{\partial \rho_{B}}{\partial t}}+\gamma^{2}v \rho_{B}\Bigg({\frac{\partial v}
{\partial t}}+ \vec{v}\cdot \vec{\nabla} v\Bigg)+\vec{\nabla} \cdot (\rho_{B}\vec{v})=0
\label{rhobcons}
\end{equation}
where $\gamma=(1-v^{2})^{-1/2}$ is the Lorentz factor.  In this work we employ the natural units $\hbar=1$, $c=1$.

Recently \cite{nos2011} we have obtained an EOS for the strongly interacting quark gluon plasma (sQGP) at zero temperature.  We performed a
gluon field separation in ``soft'' and ``hard'' components, which correspond to low and high momentum components, respectively.  In this
approach the soft gluon fields are replaced by the in-medium gluon condensates. The hard gluon fields are treated in a mean field approximation
and contribute with derivative terms in the equations of motion. Such equations solved properly may provide the time and space dependence of the
quark (or baryon) density \cite{nos2011,nos2011usa}.

Due to  the chiral phase transition, it is natural to assume that the quarks are massless and hence the system is highly relativistic.
In relativistic theories, perturbations in pressure can propagate also  in systems of massless particles. In the  Appendix, starting
from the equations of relativistic  hydrodynamics,  we derive a wave equation  for a  perturbation in the pressure, i.e., an  equation for
an acoustic wave.

The energy density is  given by \cite{nos2011,nos2011usa}:
$$
\varepsilon=\bigg({\frac{27g^{2}}{16{m_{G}}^{2}}}\bigg)  {\rho_{B}}^{2}+
\bigg({\frac{27g^{2}}{16{m_{G}}^{4}}}\bigg)
\rho_{B}  {\vec{\nabla}}^{2}\rho_{B}
+\bigg({\frac{27g^{2}}{16{m_{G}}^{6}}}\bigg)  \rho_{B}  {\vec{\nabla}}^{2}({\vec{\nabla}}^{2}\rho_{B})
$$
\begin{equation}
+\bigg({\frac{27g^{2}}{16{m_{G}}^{8}}}\bigg)  {\vec{\nabla}}^{2}\rho_{B} {\vec{\nabla}}^{2}({\vec{\nabla}}^{2}\rho_{B})
+\mathcal{B}_{QCD}
+3{\frac{\gamma_{Q}}{2{\pi}^{2}}}{\frac{{k_{F}}^{4}}{4}}
\label{eps}
\end{equation}
and the pressure is:
$$
p=\bigg({\frac{27g^{2}}{16{m_{G}}^{2}}}\bigg) \ {\rho_{B}}^{2}+\bigg({\frac{9g^{2}}{4{m_{G}}^{4}}}\bigg) \
{\rho_{B}} \ {\vec{\nabla}}^{2}{\rho_{B}}
-\bigg({\frac{9g^{2}}{8{m_{G}}^{6}}}\bigg) \ {\rho_{B}} \ {\vec{\nabla}}^{2}({\vec{\nabla}}^{2}{\rho_{B}})
$$
$$
-\bigg({\frac{9g^{2}}{16{m_{G}}^{4}}}\bigg) \vec{\nabla}{\rho_{B}} \cdot \vec{\nabla}{\rho_{B}}
+\bigg({\frac{9g^{2}}{16{m_{G}}^{6}}}\bigg) \ {\vec{\nabla}}^{2}{\rho_{B}} \ {\vec{\nabla}}^{2}{\rho_{B}}
-\bigg({\frac{9g^{2}}{8{m_{G}}^{8}}}\bigg) \ {\vec{\nabla}}^{2}{\rho_{B}} \ {\vec{\nabla}}^{2}({\vec{\nabla}}^{2}{\rho_{B}})
$$
$$
-\bigg({\frac{9g^{2}}{16{m_{G}}^{8}}}\bigg) \vec{\nabla}({\vec{\nabla}}^{2}{\rho_{B}})
\cdot \vec{\nabla}({\vec{\nabla}}^{2}{\rho_{B}})
-\bigg({\frac{9g^{2}}{8{m_{G}}^{6}}}\bigg) \vec{\nabla}{\rho_{B}} \cdot \vec{\nabla}({\vec{\nabla}}^{2}{\rho_{B}})
$$
\begin{equation}
-\mathcal{B}_{QCD}
+{\frac{\gamma_{Q}}{2{\pi}^{2}}}{\frac{{k_{F}}^{4}}{4}}
\label{pres}
\end{equation}

In (\ref{eps}) and (\ref{pres}) $\gamma_{Q}$ is the quark degeneracy factor
$\gamma_{Q} = 2 (\mbox{spin}) \times 3 (\mbox{flavor}) \, =6 $ and
$k_{F}$ is the Fermi momentum defined by the baryon number density:
\begin{equation}
\rho_{B}={\frac{\gamma_{Q}}{6{\pi}^{2}}}{{k_{F}}}^{3}
\label{kf}
\end{equation}

The other parameters $g$, $m_{G}$ and $\mathcal{B}_{QCD}$ are
the coupling of the hard gluons, the dynamical gluon mass and the bag constant in terms of the gluon condensate,
respectively.

Inserting  (\ref{eps}) and (\ref{pres})  into (1)  we  can see that   higher order derivatives in $\rho_B$ will appear.
As it will be seen, the terms with these  derivatives will generate the dispersive  terms in the final KP equations.
The  terms with derivatives in (\ref{eps}) and (\ref{pres}) exist because of the coupling  (through the coupling constant
$g$) between the quarks and the  massive gluons ($m_G$).
As explained in detail in Ref. \cite{nos2011a}, the gluon field is coupled to the quark baryon density through a Klein-Gordon
equation of motion with a source term. The gluons can thus be eliminated in favor of the quark (or baryon) density,
their inhomogeneities (expressed by non-vanishing  Laplacians) are tranferred to the quarks and terms proportional
to $\vec{\nabla}^2 \rho_B$ appear. In short: dispersion comes ultimately from the interaction between quarks and gluons and
their inhomogeneous distribution in  space.

\section{The KP equation}

We now combine the equations (\ref{eul}) and (\ref{rhobcons}) to obtain the KP equation which governs the space-time evolution of the perturbation
in the baryon density using the EOS given by (\ref{eps}) and (\ref{pres}).  As mentioned in the introduction, we will use the  RPM to obtain the
nonlinear wave equations \cite{taniuti90}.  Essentially, this formalism consists in expanding both (\ref{eul}) and (\ref{rhobcons}) in powers of a small
parameter $\sigma$.  In the following subsections we present the  application of this formalism to relativistic hydrodynamics.

We start with the cylindrical KP (cKP). Similar radially expanding perturbations have been studied in one of our previous works \cite{nos2012}
in a simplified two-dimensional  approach and with a simpler equation of state.

\subsection{Three-dimensional cylindrical coordinates}

The field velocity of the relativistic fluid is:
$$
\vec{v}=\vec{v}(r,\varphi,z,t)=\vec{v_{r}}(r,\varphi,z,t)+ \vec{v_{\varphi}}(r,\varphi,z,t)
+\vec{v_{z}}(r,\varphi,z,t)
$$ and so
$|\vec{v}|=\sqrt{{v_{r}}^{2}+{v_{\varphi}}^{2}+{v_{z}}^{2}}$ . We rewrite the equations  (\ref{eul}) and (\ref{rhobcons}) in dimensionless variables.
The perturbations in baryon density occur upon a background of density $\rho_{0}$ (the reference baryon density). It is convenient to write
the baryon density as the dimensionless quantity:
\begin{equation}
\hat{\rho}(r,\varphi,z,t)={\frac{\rho_{B}(r,\varphi,z,t)}{\rho_{0}}}
\label{vadimaftaaagain}
\end{equation}
and similarly the velocity field as:
\begin{equation}
\hat v={\frac{v}{c_{s}}}
\label{vadimaftaggeaagain}
\end{equation}
where $c_{s}$ is the speed of sound.
The components of the velocity are:
$$
\hat v_{r}(r,\varphi,z,t)={\frac{v_{r}(r,\varphi,z,t)}{c_{s}}}  \hspace{0.2cm}, \hspace{0.5cm}
\hat v_{\varphi}(r,\varphi,z,t)={\frac{v_{\varphi}(r,\varphi,z,t)}{c_{s}}}
$$
and
\begin{equation}
\hat v_{z}(r,\varphi,z,t)={\frac{v_{z}(r,\varphi,z,t)}{c_{s}}}
\label{vadimaftagaaagain}
\end{equation}

We next change variables from the space $(r,\varphi,z,t)$ to the space $(R,\Phi,Z,T)$ using the ``stretched coordinates'':
\begin{equation}
R={\frac{\sigma^{1/2}}{L}} (r-{c_{s}}t)
\hspace{0.2cm}, \hspace{0.5cm}
\Phi={\sigma^{-1/2}}\varphi
\hspace{0.2cm}, \hspace{0.5cm}
Z={\frac{\sigma}{L}}z
\hspace{0.2cm}, \hspace{0.5cm}
T={\frac{{\sigma^{3/2}}}{L}}c_{s}t
\label{stretaaagain}
\end{equation}
where $L$ is a typical scale of the problem  which renders the stretched coordinates dimensionless.
As it will be seen, the final wave equations in the three-dimensional cylindrical or cartesian coordinates
do not depend on $L$.

The next step is the expansion of the dimensionless variables in powers of the small parameter $\sigma$:
\begin{equation}
\hat\rho=1+\sigma \rho_{1}+ \sigma^{2} \rho_{2} + \sigma^{3} \rho_{3}+ \dots
\label{roexpargain}
\end{equation}
\begin{equation}
\hat {v_{r}}=\sigma v_{{r}_1}+ \sigma^{2} v_{{r}_2} + \sigma^{3} v_{{r}_3}+ \dots
\label{vexpargain}
\end{equation}
\begin{equation}
\hat {v_{\varphi}}=\sigma^{3/2} v_{{\varphi}_1}+ \sigma^{5/2} v_{{\varphi}_2}+ \sigma^{7/2} v_{{\varphi}_3}+ \dots
\label{phivexpaargain}
\end{equation}
\begin{equation}
\hat {v_{z}}=\sigma^{3/2} v_{{z}_1}+ \sigma^{5/2} v_{{z}_2}+ \sigma^{7/2} v_{{z}_3}+ \dots
\label{vexpaargain}
\end{equation}
\begin{equation}
{\hat\rho}\,^{4/3}=\big[1+(\sigma \rho_{1}+ \sigma^{2} \rho_{2}+ \dots)\big]^{4/3}
\cong 1+{\frac{4}{3}}\sigma\rho_{1}+{\frac{4}{3}} \sigma^{2} \rho_{2}+ \dots
\label{roexpaqtrgain}
\end{equation}
\begin{equation}
{\hat\rho}\,^{1/3}=\big[1+(\sigma \rho_{1}+ \sigma^{2} \rho_{2}+ \dots)\big]^{1/3}
\cong 1+{\frac{1}{3}}\sigma\rho_{1}+{\frac{1}{3}} \sigma^{2} \rho_{2}+ \dots
\label{roexpautrgain}
\end{equation}

Finally we neglect terms proportional to $\sigma^{n}$ for $n > 2$ and organize the
equations as series in powers of $\sigma$, $\sigma^{3/2}$ and $\sigma^{2}$.

From the Euler equation (\ref{eul})
we find for the radial component:
$$
\sigma \Bigg\lbrace -\bigg[\bigg({\frac{27g^{2}\,{\rho_{0}}^{2}}{8{m_{G}}^{2}}}\bigg){c_{s}}^{2}
+3\pi^{2/3}{\rho_{0}}^{4/3}{c_{s}}^{2}\bigg]
{\frac{\partial v_{{r}_1}}{\partial R}}+\bigg[\bigg({\frac{27g^{2}\,{\rho_{0}}^{2}}{8{m_{G}}^{2}}}\bigg)
+\pi^{2/3}{\rho_{0}}^{4/3}\bigg]
{\frac{\partial\rho_{1}}{\partial R}} \Bigg\rbrace
$$
$$
+\sigma^{2}\Bigg\lbrace \bigg[\bigg({\frac{27g^{2}\,{\rho_{0}}^{2}}{8{m_{G}}^{2}}}\bigg)
+\pi^{2/3}{\rho_{0}}^{4/3}\bigg]{\frac{\partial\rho_{2}}{\partial R}}
-\bigg[\bigg({\frac{27g^{2}\,{\rho_{0}}^{2}}{8{m_{G}}^{2}}}\bigg){c_{s}}^{2}
+3\pi^{2/3}{\rho_{0}}^{4/3}{c_{s}}^{2}\bigg]{\frac{\partial v_{{r}_2}}{\partial R}}
$$
$$
+\bigg[\bigg({\frac{27g^{2}\,{\rho_{0}}^{2}}{8{m_{G}}^{2}}}\bigg){c_{s}}^{2}
+3\pi^{2/3}{\rho_{0}}^{4/3}{c_{s}}^{2}\bigg]\bigg({\frac{\partial v_{{r}_1}}{\partial T}}
+v_{{r}_1}{\frac{\partial v_{{r}_1}}{\partial R}}\bigg)
$$
$$
+\bigg({\frac{27g^{2}\,{\rho_{0}}^{2}}{8{m_{G}}^{2}}}\bigg)\rho_{1}{\frac{\partial \rho_{1}}{\partial R}}
+\pi^{2/3}{\rho_{0}}^{4/3}{\frac{\rho_{1}}{3}}{\frac{\partial \rho_{1}}{\partial R}}
$$
$$
-\bigg[\bigg({\frac{27g^{2}\,{\rho_{0}}^{2}}{8{m_{G}}^{2}}}\bigg)2{c_{s}}^{2}
+4\pi^{2/3}{\rho_{0}}^{4/3}{c_{s}}^{2}\bigg]\rho_{1}{\frac{\partial v_{{r}_1}}{\partial R}}
$$
\begin{equation}
-\bigg[\bigg({\frac{27g^{2}\,{\rho_{0}}^{2}}{8{m_{G}}^{2}}}\bigg){c_{s}}^{2}
+\pi^{2/3}{\rho_{0}}^{4/3}{c_{s}}^{2}\bigg]v_{{r}_1}{\frac{\partial \rho_{1}}{\partial R}}
+\bigg({\frac{9g^{2}\,{\rho_{0}}^{2}}{4{m_{G}}^{4}L^{2}}}\bigg)
{\frac{\partial^{3} \rho_{1}}{\partial R^{3}}} \Bigg\rbrace =0
\label{eulerxexpprgain}
\end{equation}
For the angular component:
$$
\sigma^{3/2} \Bigg\lbrace -\bigg[\bigg({\frac{27g^{2}\,{\rho_{0}}^{2}}{8{m_{G}}^{2}}}\bigg){c_{s}}^{2}
+3\pi^{2/3}{\rho_{0}}^{4/3}{c_{s}}^{2}\bigg]
{\frac{\partial v_{{\varphi}_1}}{\partial R}}
$$
\begin{equation}
+\bigg[\bigg({\frac{27g^{2}\,{\rho_{0}}^{2}}{8{m_{G}}^{2}}}\bigg)
+\pi^{2/3}{\rho_{0}}^{4/3}\bigg]{\frac{1}{T}}
{\frac{\partial\rho_{1}}{\partial \Phi}} \Bigg\rbrace = 0
\label{euleryexprpzgainphi}
\end{equation}
and for the component in the $z$ direction:
\begin{equation}
\sigma^{3/2} \Bigg\lbrace -\bigg[\bigg({\frac{27g^{2}\,{\rho_{0}}^{2}}{8{m_{G}}^{2}}}\bigg){c_{s}}^{2}
+3\pi^{2/3}{\rho_{0}}^{4/3}{c_{s}}^{2}\bigg]
{\frac{\partial v_{{z}_1}}{\partial R}}+\bigg[\bigg({\frac{27g^{2}\,{\rho_{0}}^{2}}{8{m_{G}}^{2}}}\bigg)
+\pi^{2/3}{\rho_{0}}^{4/3}\bigg]
{\frac{\partial\rho_{1}}{\partial Z}} \Bigg\rbrace = 0
\label{euleryexprpzgain}
\end{equation}
Performing the same calculations for the continuity equation (\ref{rhobcons})
we find:
$$
\sigma \Bigg\lbrace  {\frac{\partial v_{{r}_1}}{\partial R}}
- {\frac{\partial \rho_{1}}{\partial R}}   \Bigg\rbrace+
\sigma^{2} \Bigg\lbrace {\frac{\partial v_{{r}_2}}{\partial R}}
-{\frac{\partial\rho_{2}}{\partial R}}
+{\frac{\partial\rho_{1}}{\partial T}}
$$
\begin{equation}
+\rho_{1}{\frac{\partial v_{{r}_1}}{\partial R}}+v_{{r}_1}{\frac{\partial \rho_{1}}{\partial R}}
-{c_{s}}^{2}v_{{r}_1}{\frac{\partial v_{{r}_1}}{\partial R}}+{\frac{v_{{r}_1}}{T}}+
{\frac{\partial v_{{z}_1}}{\partial Z}}+{\frac{1}{T}}
{\frac{\partial v_{{\varphi}_1}}{\partial \Phi}} \Bigg\rbrace =0
\label{contexprpcgainn}
\end{equation}
In the last four equations each bracket must vanish independently and so
$\lbrace \dots  \rbrace = 0$.
From  the terms proportional to $\sigma$ we
obtain the identity:
\begin{equation}
\bigg({\frac{27g^{2}\,{\rho_{0}}^{2}}{8{m_{G}}^{2}}}\bigg){c_{s}}^{2}
+3\pi^{2/3}{\rho_{0}}^{4/3}{c_{s}}^{2}=\bigg({\frac{27g^{2}\,{\rho_{0}}^{2}}{8{m_{G}}^{2}}}\bigg)
+\pi^{2/3}{\rho_{0}}^{4/3} = A
\label{aconsaggainn}
\end{equation}
which defines the constant $A$ and from which we obtain the speed of sound for a given background density ${\rho_{0}}$:
\begin{equation}
{c_{s}}^{2}={\frac{\bigg({\frac{27g^{2}\,{\rho_{0}}^{2}}{8{m_{G}}^{2}}}\bigg)+\pi^{2/3}{\rho_{0}}^{4/3}}
{\bigg({\frac{27g^{2}\,{\rho_{0}}^{2}}{8{m_{G}}^{2}}}\bigg)+3\pi^{2/3}{\rho_{0}}^{4/3}}}
\label{csrelationaggainn}
\end{equation}
and also
\begin{equation}
\rho_{1}=v_{{r}_1}
\label{rhoumr}
\end{equation}
From the terms proportional to $\sigma^{3/2}$ using the $A$ constant we find:
\begin{equation}
{\frac{\partial v_{{\varphi}_1}}{\partial R}}={\frac{1}{T}}{\frac{\partial\rho_{1}}{\partial \Phi}}
\label{tdum}
\end{equation}
and
\begin{equation}
{\frac{\partial v_{{z}_1}}{\partial R}}={\frac{\partial\rho_{1}}{\partial Z}}
\label{tddois}
\end{equation}

Inserting the results (\ref{aconsaggainn}), (\ref{rhoumr}), (\ref{tdum}) and (\ref{tddois}) into the terms
proportional to $\sigma^{2}$ in (\ref{eulerxexpprgain}) and (\ref{contexprpcgainn}), we find after some algebra, the cylindrical
Kadomtsev-Petviashvili (cKP) equation \cite{mushtaq}:
$$
{\frac{\partial}{\partial R}}\Bigg\lbrace{\frac{\partial\rho_{1}}{\partial T}}
+\bigg[{\frac{(2-{c_{s}}^{2})}{2}}-\bigg({\frac{27g^{2}\,{\rho_{0}}^{2}}{8{m_{G}}^{2}}}\bigg)
{\frac{(2{c_{s}}^{2}-1)}{2A}}-{\frac{\pi^{2/3}{\rho_{0}}^{4/3}}{A}}\bigg({c_{s}}^{2}-{\frac{1}{6}}\bigg)\bigg]
\rho_{1}{\frac{\partial \rho_{1}}{\partial R}}
$$
\begin{equation}
+\bigg[{\frac{9g^{2}\,{\rho_{0}}^{2}}{8{m_{G}}^{4}L^{2}A}}\bigg]
{\frac{\partial^{3} \rho_{1}}{\partial R^{3}}}+{\frac{\rho_{1}}{2T}}\Bigg\rbrace +
{\frac{1}{2T^{2}}}
{\frac{\partial^{2} \rho_{1}}{\partial \Phi^{2}}}+
{\frac{1}{2}}{\frac{\partial^{2}\rho_{1}}{\partial Z^{2}}}=0
\label{kpxxyyttaggainnalm}
\end{equation}
From the second identity of (\ref{aconsaggainn}) we may write
\begin{equation}
\bigg({\frac{27g^{2}\,{\rho_{0}}^{2}}{8{m_{G}}^{2}}}\bigg)=A-\pi^{2/3}{\rho_{0}}^{4/3}
\label{againaused}
\end{equation}
Inserting (\ref{againaused}) in the coefficient of the nonlinear term in (\ref{kpxxyyttaggainnalm}) the cKP becomes:
$$
{\frac{\partial}{\partial R}}\Bigg\lbrace{\frac{\partial\rho_{1}}{\partial T}}
+\bigg[{\frac{3}{2}}(1-{c_{s}}^{2})-{\frac{\pi^{2/3}{\rho_{0}}^{4/3}}{3A}}\bigg]
\rho_{1}{\frac{\partial \rho_{1}}{\partial R}}
+\bigg[{\frac{9g^{2}\,{\rho_{0}}^{2}}{8{m_{G}}^{4}L^{2}A}}\bigg]
{\frac{\partial^{3} \rho_{1}}{\partial R^{3}}}
+{\frac{\rho_{1}}{2T}}\Bigg\rbrace
$$
\begin{equation}
+{\frac{1}{2T^{2}}}
{\frac{\partial^{2} \rho_{1}}{\partial \Phi^{2}}}+
{\frac{1}{2}}{\frac{\partial^{2}\rho_{1}}{\partial Z^{2}}}=0
\label{kpxxyyttaggainna}
\end{equation}
Returning this cKP equation to the three dimension cylindrical space yields:
$$
{\frac{\partial}{\partial r}}\Bigg\lbrace{\frac{\partial\hat\rho_{1}}{\partial t}}
+{c_{s}}{\frac{\partial\hat\rho_{1}}{\partial r}}
+\bigg[{\frac{3}{2}}(1-{c_{s}}^{2})-{\frac{\pi^{2/3}{\rho_{0}}^{4/3}}{3A}}\bigg]{c_{s}}
\hat\rho_{1}{\frac{\partial\hat\rho_{1}}{\partial r}}
+\bigg[{\frac{9g^{2}\,{\rho_{0}}^{2}c_{s}}{8{m_{G}}^{4}A}}\bigg]
{\frac{\partial^{3}\hat\rho_{1}}{\partial r^{3}}}+
{\frac{\hat\rho_{1}}{2t}}\Bigg\rbrace
$$
\begin{equation}
+{\frac{1}{2c_{s}t^{2}}}{\frac{\partial^{2}\hat\rho_{1}}{\partial \varphi^{2}}}
+{\frac{c_{s}}{2}}{\frac{\partial^{2}\hat\rho_{1}}{\partial z^{2}}}=0
\label{r3dckp}
\end{equation}
which is the cKP equation for the second term
of the expansion (\ref{roexpargain}),  the small perturbation given by $\hat\rho_{1}\equiv \sigma\rho_{1}$.

\subsection{Three-dimensional cartesian coordinates}

We now write the field velocity of the relativistic fluid as:
$$\vec{v}=\vec{v}(x,y,t)=\vec{v_{x}}(x,y,t)+\vec{v_{y}}(x,y,t)+\vec{v_{z}}(x,y,t)$$ and so
$|\vec{v}|=\sqrt{{v_{x}}^{2}+{v_{y}}^{2}+{v_{z}}^{2}}$ .

We follow the same steps described in the cylindrical case to obtain the KP equation:

$1)$ Rewrite the equations (\ref{eul}) and (\ref{rhobcons}) in dimensionless variables:
\begin{equation}
\hat{\rho}(x,y,z,t)={\frac{\rho_{B}(x,y,z,t)}{\rho_{0}}}
\label{vadimaft}
\end{equation}
\begin{equation}
\hat v={\frac{v}{c_{s}}}
\label{vadimaftagge}
\end{equation}
The components of the velocity are given by:
$$
\hat v_{x}(x,y,z,t)={\frac{v_{x}(x,y,z,t)}{c_{s}}}  \hspace{0.2cm}, \hspace{0.5cm} \hat v_{y}(x,y,z,t)={\frac{v_{y}(x,y,z,t)}{c_{s}}}
$$
and
\begin{equation}
\hat v_{z}(r,\varphi,z,t)={\frac{v_{z}(r,\varphi,z,t)}{c_{s}}}
\label{vadimaftag}
\end{equation}
$2)$ Transform the equations (\ref{eul}) and (\ref{rhobcons}) (now in dimensionless variables)
from the space $(x,y,z,t)$ to the space $(X,Y,Z,T)$ using the ``stretched coordinates'':
\begin{equation}
X={\frac{\sigma^{1/2}}{L}} (x-{c_{s}}t)
\hspace{0.2cm}, \hspace{0.5cm}
Y={\frac{\sigma}{L}}y
\hspace{0.2cm}, \hspace{0.5cm}
Z={\frac{\sigma}{L}}z
\hspace{0.2cm}, \hspace{0.5cm}
T={\frac{{\sigma^{3/2}}}{L}}c_{s}t
\label{streta}
\end{equation}
$3)$ Perform the expansions of the dimensionless variables:
\begin{equation}
\hat\rho=1+\sigma \rho_{1}+ \sigma^{2} \rho_{2} + \sigma^{3} \rho_{3}+ \dots
\label{roexpa}
\end{equation}
\begin{equation}
\hat {v_{x}}=\sigma v_{{x}_1}+ \sigma^{2} v_{{x}_2} + \sigma^{3} v_{{x}_3}+ \dots
\label{vexpa}
\end{equation}
\begin{equation}
\hat {v_{y}}=\sigma^{3/2} v_{{y}_1}+ \sigma^{2} v_{{y}_2}+ \sigma^{5/2} v_{{y}_3}+ \dots
\label{vexpaa}
\end{equation}
\begin{equation}
\hat {v_{z}}=\sigma^{3/2} v_{{z}_1}+ \sigma^{2} v_{{z}_2}+ \sigma^{5/2} v_{{z}_3}+ \dots
\label{vexpaa}
\end{equation}
\begin{equation}
{\hat\rho}\,^{4/3}
\cong 1+{\frac{4}{3}}\sigma\rho_{1}+{\frac{4}{3}} \sigma^{2} \rho_{2}+ \dots
\label{roexpaqt}
\end{equation}
\begin{equation}
{\hat\rho}\,^{1/3}
\cong 1+{\frac{1}{3}}\sigma\rho_{1}+{\frac{1}{3}} \sigma^{2} \rho_{2}+ \dots
\label{roexpaut}
\end{equation}
$4)$ Neglect terms proportional to $\sigma^{n}$ for $n > 2$ and organize the
equations as series in powers of $\sigma$, $\sigma^{3/2}$ and $\sigma^{2}$.

After these manipulations the $x$, $y$ and $z$ components of the Euler equation become:
$$
\sigma \Bigg\lbrace -\bigg[\bigg({\frac{27g^{2}\,{\rho_{0}}^{2}}{8{m_{G}}^{2}}}\bigg){c_{s}}^{2}
+3\pi^{2/3}{\rho_{0}}^{4/3}{c_{s}}^{2}\bigg]
{\frac{\partial v_{{x}_1}}{\partial X}}+\bigg[\bigg({\frac{27g^{2}\,{\rho_{0}}^{2}}{8{m_{G}}^{2}}}\bigg)
+\pi^{2/3}{\rho_{0}}^{4/3}\bigg]
{\frac{\partial\rho_{1}}{\partial X}} \Bigg\rbrace
$$
$$
+\sigma^{2}\Bigg\lbrace \bigg[\bigg({\frac{27g^{2}\,{\rho_{0}}^{2}}{8{m_{G}}^{2}}}\bigg)
+\pi^{2/3}{\rho_{0}}^{4/3}\bigg]{\frac{\partial\rho_{2}}{\partial X}}
-\bigg[\bigg({\frac{27g^{2}\,{\rho_{0}}^{2}}{8{m_{G}}^{2}}}\bigg){c_{s}}^{2}
+3\pi^{2/3}{\rho_{0}}^{4/3}{c_{s}}^{2}\bigg]{\frac{\partial v_{{x}_2}}{\partial X}}
$$
$$
+\bigg[\bigg({\frac{27g^{2}\,{\rho_{0}}^{2}}{8{m_{G}}^{2}}}\bigg){c_{s}}^{2}
+3\pi^{2/3}{\rho_{0}}^{4/3}{c_{s}}^{2}\bigg]\bigg({\frac{\partial v_{{x}_1}}{\partial T}}
+v_{{x}_1}{\frac{\partial v_{{x}_1}}{\partial X}}\bigg)
$$
$$
+\bigg({\frac{27g^{2}\,{\rho_{0}}^{2}}{8{m_{G}}^{2}}}\bigg)\rho_{1}{\frac{\partial \rho_{1}}{\partial X}}
+\pi^{2/3}{\rho_{0}}^{4/3}{\frac{\rho_{1}}{3}}{\frac{\partial \rho_{1}}{\partial X}}
$$
$$
-\bigg[\bigg({\frac{27g^{2}\,{\rho_{0}}^{2}}{8{m_{G}}^{2}}}\bigg)2{c_{s}}^{2}
+4\pi^{2/3}{\rho_{0}}^{4/3}{c_{s}}^{2}\bigg]\rho_{1}{\frac{\partial v_{{x}_1}}{\partial X}}
$$
\begin{equation}
-\bigg[\bigg({\frac{27g^{2}\,{\rho_{0}}^{2}}{8{m_{G}}^{2}}}\bigg){c_{s}}^{2}
+\pi^{2/3}{\rho_{0}}^{4/3}{c_{s}}^{2}\bigg]v_{{x}_1}{\frac{\partial \rho_{1}}{\partial X}}
+\bigg({\frac{9g^{2}\,{\rho_{0}}^{2}}{4{m_{G}}^{4}L^{2}}}\bigg)
{\frac{\partial^{3} \rho_{1}}{\partial X^{3}}} \Bigg\rbrace =0
\label{eulerxexpx}
\end{equation}
$$
\sigma^{3/2} \Bigg\lbrace -\bigg[\bigg({\frac{27g^{2}\,{\rho_{0}}^{2}}{8{m_{G}}^{2}}}\bigg){c_{s}}^{2}
+3\pi^{2/3}{\rho_{0}}^{4/3}{c_{s}}^{2}\bigg]
{\frac{\partial v_{{y}_1}}{\partial X}}+\bigg[\bigg({\frac{27g^{2}\,{\rho_{0}}^{2}}{8{m_{G}}^{2}}}\bigg)
+\pi^{2/3}{\rho_{0}}^{4/3}\bigg]
{\frac{\partial\rho_{1}}{\partial Y}} \Bigg\rbrace
$$
\begin{equation}
+\sigma^{2} \Bigg\lbrace -\bigg[\bigg({\frac{27g^{2}\,{\rho_{0}}^{2}}{8{m_{G}}^{2}}}\bigg){c_{s}}^{2}
+3\pi^{2/3}{\rho_{0}}^{4/3}{c_{s}}^{2}\bigg]
{\frac{\partial v_{{y}_2}}{\partial X}}\Bigg\rbrace = 0
\label{eulerexpy}
\end{equation}
and
$$
\sigma^{3/2} \Bigg\lbrace -\bigg[\bigg({\frac{27g^{2}\,{\rho_{0}}^{2}}{8{m_{G}}^{2}}}\bigg){c_{s}}^{2}
+3\pi^{2/3}{\rho_{0}}^{4/3}{c_{s}}^{2}\bigg]
{\frac{\partial v_{{z}_1}}{\partial X}}+\bigg[\bigg({\frac{27g^{2}\,{\rho_{0}}^{2}}{8{m_{G}}^{2}}}\bigg)
+\pi^{2/3}{\rho_{0}}^{4/3}\bigg]
{\frac{\partial\rho_{1}}{\partial Z}} \Bigg\rbrace
$$
\begin{equation}
+\sigma^{2} \Bigg\lbrace -\bigg[\bigg({\frac{27g^{2}\,{\rho_{0}}^{2}}{8{m_{G}}^{2}}}\bigg){c_{s}}^{2}
+3\pi^{2/3}{\rho_{0}}^{4/3}{c_{s}}^{2}\bigg]
{\frac{\partial v_{{z}_2}}{\partial X}}\Bigg\rbrace = 0
\label{eulerexpz}
\end{equation}
For the continuity equation we obtain:
$$
\sigma \Bigg\lbrace  {\frac{\partial v_{{x}_1}}{\partial X}}
- {\frac{\partial \rho_{1}}{\partial X}}   \Bigg\rbrace
$$
\begin{equation}
+\sigma^{2} \Bigg\lbrace {\frac{\partial v_{{x}_2}}{\partial X}}
-{\frac{\partial\rho_{2}}{\partial X}}
+{\frac{\partial\rho_{1}}{\partial T}}
+\rho_{1}{\frac{\partial v_{{x}_1}}{\partial X}}+v_{{x}_1}{\frac{\partial \rho_{1}}{\partial X}}
-{c_{s}}^{2}v_{{x}_1}{\frac{\partial v_{{x}_1}}{\partial X}}+{\frac{\partial v_{{y}_1}}{\partial Y}}+{\frac{\partial v_{{z}_1}}{\partial Y}} \Bigg\rbrace =0
\label{contexpxy}
\end{equation}
Again, in the last four equations each bracket must vanish independently.
From  the terms proportional to $\sigma$ we
obtain the same $A$ constant as in the cylindrical case given by (\ref{aconsaggainn}), the same expression for the speed of sound (\ref{csrelationaggainn}) and
\begin{equation}
\rho_{1}=v_{{x}_1}
\label{rhoum}
\end{equation}
From the terms proportional to $\sigma^{3/2}$ we find
\begin{equation}
{\frac{\partial v_{{y}_1}}{\partial X}}={\frac{\partial\rho_{1}}{\partial Y}}
\label{rhoumderi}
\end{equation}
and
\begin{equation}
{\frac{\partial v_{{z}_1}}{\partial X}}={\frac{\partial\rho_{1}}{\partial Z}}
\label{rhoumderiother}
\end{equation}
In (\ref{eulerexpy}) and (\ref{eulerexpz}) we have from the terms proportional to $\sigma^{2}$:
\begin{equation}
{\frac{\partial v_{{y}_2}}{\partial X}}={\frac{\partial v_{{z}_2}}{\partial X}}=0
\label{rhodoisderi}
\end{equation}
Inserting the results (\ref{aconsaggainn}), (\ref{rhoum}), (\ref{rhoumderi}), (\ref{rhoumderiother}) and (\ref{rhodoisderi}) into the terms
proportional to $\sigma^{2}$ in (\ref{eulerxexpx}) and (\ref{contexpxy}), we find after some algebra, the
Kadomtsev-Petviashvili (KP) equation \cite{kp2004,kp2010}:
$$
{\frac{\partial}{\partial X}}\Bigg\lbrace{\frac{\partial\rho_{1}}{\partial T}}
+\bigg[{\frac{(2-{c_{s}}^{2})}{2}}-\bigg({\frac{27g^{2}\,{\rho_{0}}^{2}}{8{m_{G}}^{2}}}\bigg)
{\frac{(2{c_{s}}^{2}-1)}{2A}}-{\frac{\pi^{2/3}{\rho_{0}}^{4/3}}{A}}\bigg({c_{s}}^{2}-{\frac{1}{6}}\bigg)\bigg]
\rho_{1}{\frac{\partial \rho_{1}}{\partial X}}
$$
\begin{equation}
+\bigg[{\frac{9g^{2}\,{\rho_{0}}^{2}}{8{m_{G}}^{4}L^{2}A}}\bigg]
{\frac{\partial^{3} \rho_{1}}{\partial X^{3}}}\Bigg\rbrace +
{\frac{1}{2}}{\frac{\partial^{2}\rho_{1}}{\partial Y^{2}}}+
{\frac{1}{2}}{\frac{\partial^{2}\rho_{1}}{\partial Z^{2}}}=0
\label{kpxxyytt}
\end{equation}
Inserting (\ref{againaused}) in (\ref{kpxxyytt}), the KP with simplified coefficient
for the nonlinear term is given by:
\begin{equation}
{\frac{\partial}{\partial X}}\Bigg\lbrace{\frac{\partial\rho_{1}}{\partial T}}+
\bigg[{\frac{3}{2}}(1-{c_{s}}^{2})-{\frac{\pi^{2/3}{\rho_{0}}^{4/3}}{3A}}\bigg]
\rho_{1}{\frac{\partial \rho_{1}}{\partial X}}
+\bigg[{\frac{9g^{2}\,{\rho_{0}}^{2}}{8{m_{G}}^{4}L^{2}A}}\bigg]
{\frac{\partial^{3} \rho_{1}}{\partial X^{3}}}\Bigg\rbrace
+{\frac{1}{2}}{\frac{\partial^{2}\rho_{1}}{\partial Y^{2}}}+
{\frac{1}{2}}{\frac{\partial^{2}\rho_{1}}{\partial Z^{2}}}=0
\label{kpxxyyttsimp}
\end{equation}
Rewriting this KP equation back in the three dimensional cartesian space we find:
\begin{equation}
{\frac{\partial}{\partial x}}\Bigg\lbrace{\frac{\partial\hat\rho_{1}}{\partial t}}
+{c_{s}}{\frac{\partial\hat\rho_{1}}{\partial x}}
+\bigg[{\frac{3}{2}}(1-{c_{s}}^{2})-{\frac{\pi^{2/3}{\rho_{0}}^{4/3}}{3A}}\bigg]{c_{s}}
\hat\rho_{1}{\frac{\partial\hat\rho_{1}}{\partial x}}
+\bigg[{\frac{9g^{2}\,{\rho_{0}}^{2}c_{s}}{8{m_{G}}^{4}A}}\bigg]
{\frac{\partial^{3}\hat\rho_{1}}{\partial x^{3}}}\Bigg\rbrace+{\frac{c_{s}}{2}}{\frac{\partial^{2}\hat\rho_{1}}{\partial y^{2}}}
+{\frac{c_{s}}{2}}{\frac{\partial^{2}\hat\rho_{1}}{\partial z^{2}}}=0
\label{kpxyztsual}
\end{equation}
which is the KP equation for the small perturbation
$\hat\rho_{1}\equiv \sigma\rho_{1}$, the second term
of the expansion (\ref{roexpa}).

The techniques employed in this section are well suited to treat problems where a long wave approximation can be made.
Having derived the relevant differential equation,  we can check whether the obtained equation is
consistent with the physical picture of a small amplitude and long wave length perturbation propagating over large distances.
We shall follow the analysis performed in Ref. \cite{leblond}.  Let us assume that the above equation has a solitary wave solution
with a typical large length $L \simeq 1/\sigma \,\,$   $ \,\, (\sigma << 1)$. The dispersion term is about
$ \frac{\partial^4 \hat\rho_{1}}{\partial x^4}  \simeq  \sigma^4 \hat\rho_{1}$. It must arise at a propagation distance (or equivalently propagation
time T) D, accounted for in the equation by the term $\frac{\partial^2 \hat\rho_{1}} {\partial x \partial t}   \simeq \sigma  \frac{\hat\rho_{1}}{T} $.
If both the dispersion and propagation terms have the same size, then $T  \simeq  D \simeq  1/\sigma^3$. Regarding the nonlinear term, if it has the form
$\frac{\partial}{\partial x}  ( \hat\rho_{1} \frac{\partial \hat\rho_{1}} { \partial x }) $ its order of magnitude is
$\hat\rho_{1}^2 \sigma^2$. The formation of the soliton requires that the nonlinear effect balances the dispersion. Hence it must have the same
order of magnitude and  $\hat\rho_{1}^2 \sigma^2 =  \hat\rho_{1} \sigma^4$. Hence $ \hat\rho_{1} \simeq  \sigma^2 $. We can then conclude that
$\hat\rho_{1} \, << \, L \, << \, D$ and the above equation describes the propagation of a wave with small amplitude  $( \hat\rho_{1})$ and large wave
length $(L)$ which travels  large distances $(D)$.  The last two terms in (\ref{kpxyztsual}) describe the transverse evolution of  the wave. We can
estimate their sizes only if we make assumptions about the transverse length scales. In most cases the resulting flow is one-dimensional along the $x$
direction with some ``leakage'' to the transverse directions. In view of these estimates, we believe that the use of the RPM in this context is justified.

\subsection{Some particular cases}

In the one dimensional cartesian relativistic fluid dynamics we have
$\vec{v}=\vec{v}(x,t)$
and $\rho_{B}=\rho_{B}(x,t)$. Repeating all the steps of the last subsection for one dimension,
the reductive perturbation method reduces to the formalism previously used in
\cite{nos2010,fn1,fn2,fn3,fn4,frsw,abu} and we find the following particular cases
of (\ref{kpxyztsual}):

$(I)$ Neglecting the $y$ and $z$ dependence, the (\ref{kpxyztsual}) becomes the Korteweg-de Vries equation (KdV) similar to the KdV found in \cite{nos2011a}:
$$
{\frac{\partial\hat\rho_{1}}{\partial t}}
+{c_{s}}{\frac{\partial\hat\rho_{1}}{\partial x}}+
\bigg[{\frac{(2-{c_{s}}^{2})}{2}}-\bigg({\frac{27g^{2}\,{\rho_{0}}^{2}}{8{m_{G}}^{2}}}\bigg)
{\frac{(2{c_{s}}^{2}-1)}{2A}}-{\frac{\pi^{2/3}{\rho_{0}}^{4/3}}{A}}\bigg({c_{s}}^{2}-{\frac{1}{6}}\bigg)\bigg]{c_{s}}
\hat\rho_{1}{\frac{\partial\hat\rho_{1}}{\partial x}}
$$
\begin{equation}
+\bigg[{\frac{9g^{2}\,{\rho_{0}}^{2}c_{s}}{8{m_{G}}^{4}A}}\bigg]
{\frac{\partial^{3}\hat\rho_{1}}{\partial x^{3}}}=0
\label{kdvxt}
\end{equation}
Taking the limit $m_{G} \, \rightarrow \, \infty$ we obtain from (\ref{aconsaggainn}) and (\ref{csrelationaggainn}):
$$
A=\pi^{2/3}{\rho_{0}}^{4/3}  \,\,\,\, , \hspace{2cm} {c_{s}}^{2}={\frac{1}{3}}
$$
and (\ref{kdvxt}) becomes:
\begin{equation}
{\frac{\partial\hat\rho_{1}}{\partial t}}+
c_{s}{\frac{\partial \hat\rho_{1}}{\partial x}}+
{\frac{2}{3}}c_{s}\hat\rho_{1}{\frac{\partial \hat\rho_{1}}{\partial x}}=0
\label{bwmitxitauXt}
\end{equation}
and we recover exactly the result found in \cite{nos2010}, the so called breaking wave
equation for $\hat\rho_{1}$ at zero temperature in the QGP with the MIT equation of state.

$(II)$ Neglecting the spatial derivatives in (\ref{eps}) and (\ref{pres}), equation
(\ref{kdvxt}) reduces to:
\begin{equation}
{\frac{\partial\hat\rho_{1}}{\partial t}}
+{c_{s}}{\frac{\partial\hat\rho_{1}}{\partial x}}+
\bigg[{\frac{(2-{c_{s}}^{2})}{2}}-\bigg({\frac{27g^{2}\,{\rho_{0}}^{2}}{8{m_{G}}^{2}}}\bigg)
{\frac{(2{c_{s}}^{2}-1)}{2A}}-{\frac{\pi^{2/3}{\rho_{0}}^{4/3}}{A}}\bigg({c_{s}}^{2}-{\frac{1}{6}}\bigg)\bigg]{c_{s}}
\hat\rho_{1}{\frac{\partial\hat\rho_{1}}{\partial x}}=0
\label{bwqxt}
\end{equation}
which is also a breaking wave equation for $\hat\rho_{1}$ with the $\rho_{0}$, $m_{G}$ and $g$ dependence in its
coefficients.

\section{Non-relativistic Limit}

The non-relativistic version of the continuity equation (\ref{eul}) is given by \cite{land,wein}:
\begin{equation}
{\frac{\partial \rho_{B}}{\partial t}} + {\vec{\nabla}} \cdot (\rho_{B} {\vec{v}})=0
\label{contnr}
\end{equation}
and for the Euler equation we have (\ref{rhobcons}) \cite{land,wein}:
\begin{equation}
{\frac{\partial \vec{v}}{\partial t}} +(\vec{v} \cdot \vec{\nabla}) \vec{v}=
-\bigg({\frac{1}{\rho}}\bigg) \vec{\nabla} p
\label{eulnralmost}
\end{equation}
where $\rho$ is the volumetric density of fluid matter.  In this work we study perturbations for
baryon density in the sQGP fluid, so we define the `` effective baryon mass $\mathcal{M}$ in sQGP '' :
\begin{equation}
\rho=\mathcal{M} \rho_{B}
\label{ebmass}
\end{equation}
which will be determined latter.
Substituting (\ref{ebmass}) in (\ref{eulnralmost}) we find the non-relativistic version for
the Euler equation in the  sQGP:
\begin{equation}
{\frac{\partial \vec{v}}{\partial t}} +(\vec{v} \cdot \vec{\nabla}) \vec{v}=
-\bigg({\frac{1}{\mathcal{M} \rho_{B}}}\bigg) \vec{\nabla} p
\label{eulnr}
\end{equation}

Performing all the calculations described in the last section for the combination of
(\ref{contnr}) and (\ref{eulnr}) we find the cKP equation in non-relativistic hydrodynamics:
$$
{\frac{\partial}{\partial r}}\Bigg\lbrace{\frac{\partial\hat\rho_{1}}{\partial t}}
+{c_{s}}{\frac{\partial\hat\rho_{1}}{\partial r}}
+\bigg[{\frac{3}{2}}-{\frac{\pi^{2/3}{\rho_{0}}^{1/3}}{3\mathcal{M}{c_{s}}^{2}}}\bigg]{c_{s}}
\hat\rho_{1}{\frac{\partial\hat\rho_{1}}{\partial r}}
+\bigg[{\frac{9g^{2}\,{\rho_{0}}}{8\mathcal{M}{{m_{G}}^{4}}{c_{s}}}}\bigg]
{\frac{\partial^{3}\hat\rho_{1}}{\partial x^{3}}}+
{\frac{\hat\rho_{1}}{2t}}\Bigg\rbrace
$$
\begin{equation}
+{\frac{1}{2c_{s}t^{2}}}{\frac{\partial^{2}\hat\rho_{1}}{\partial \varphi^{2}}}
+{\frac{c_{s}}{2}}{\frac{\partial^{2}\hat\rho_{1}}{\partial z^{2}}}=0
\label{r3dckpnr}
\end{equation}
and the KP equation in three-dimensional cartesian coordinates:
\begin{equation}
{\frac{\partial}{\partial x}}\Bigg\lbrace{\frac{\partial\hat\rho_{1}}{\partial t}}
+{c_{s}}{\frac{\partial\hat\rho_{1}}{\partial x}}
+\bigg[{\frac{3}{2}}-{\frac{\pi^{2/3}{\rho_{0}}^{1/3}}{3\mathcal{M}{c_{s}}^{2}}}\bigg]{c_{s}}
\hat\rho_{1}{\frac{\partial\hat\rho_{1}}{\partial x}}
+\bigg[{\frac{9g^{2}\,{\rho_{0}}}{8\mathcal{M}{{m_{G}}^{4}}{c_{s}}}}\bigg]
{\frac{\partial^{3}\hat\rho_{1}}{\partial x^{3}}}\Bigg\rbrace+{\frac{c_{s}}{2}}{\frac{\partial^{2}\hat\rho_{1}}{\partial y^{2}}}
+{\frac{c_{s}}{2}}{\frac{\partial^{2}\hat\rho_{1}}{\partial z^{2}}}=0
\label{kpxyztsualnr}
\end{equation}
which are the non-relativistic versions of (\ref{r3dckp}) and (\ref{kpxyztsual}) respectively.
During the derivation in both cases we find from the terms proportional to $\sigma$ in the Euler equation that:
\begin{equation}
\mathcal{M}=\bigg({\frac{27g^{2}\,{\rho_{0}}}{8{m_{G}}^{2}{c_{s}}^{2}}}\bigg)
+{\frac{\pi^{2/3}{\rho_{0}}^{1/3}}{{c_{s}}^{2}}}
\label{ebmassexp}
\end{equation}

We end this section mentioning that it is possible to obtain (\ref{r3dckpnr}) and (\ref{kpxyztsualnr}) directly from
(\ref{r3dckp}) and (\ref{kpxyztsual}) respectively, performing the two non-relativistic
approximations:
\begin{equation}
a) \hspace{3.0cm} {c_{s}}^{2} \rightarrow 0
\label{nra1}
\end{equation}
and
\begin{equation}
b) \hspace{2.5cm} A=\mathcal{M}{{\rho_{0}}{c_{s}}^{2}}
\label{nra2}
\end{equation}
where $A$ is given by (\ref{aconsaggainn}) and $\mathcal{M}$ by (\ref{ebmassexp}).

\section{Analytical  solutions}

\subsection{Soliton-like solutions}

There are several methods to solve the KP equation such as the generalized expansion method \cite{kp2010,moslempp17},
inverse scattering transform (IST) \cite{segur,novikov} and others.  The KP is also tractable by the Riemann theta functions,
 as it was shown in \cite{dubrovin}, where other solution techniques  are discussed.  In this work we are only interested
in the  particular case of the solitonic solution.

In this section we present the analytical soliton-like solution of the cKP and KP equation given by
(\ref{r3dckp}) and (\ref{kpxyztsual}) respectively. The KP equation is an integrable system in three
dimensions in the same way as the KdV is in one dimension.  We introduce a set of coordinates that transforms
(\ref{r3dckp}) in an ordinary KdV, which is a solvable equation, and we also present the analytical solution of
(\ref{kpxyztsual}).  In order to simplify the notation in equations (\ref{r3dckp}) and (\ref{kpxyztsual}) we define
the constants:
\begin{equation}
\alpha \equiv \bigg[{\frac{3}{2}}(1-{c_{s}}^{2})-{\frac{\pi^{2/3}{\rho_{0}}^{4/3}}{3A}}\bigg]c_{s}
\end{equation}
and
\begin{equation}
\beta \equiv \bigg[{\frac{9g^{2}\,{\rho_{0}}^{2}c_{s}}{8{m_{G}}^{4}A}}\bigg]
\end{equation}
In order to  solve (\ref{r3dckp})  analytically  we introduce the following coordinates \cite{jukui,yunliang,mushtaq,sahu11}:
\begin{equation}
\xi=ar+bz-d{\frac{{c_{s}}\varphi^{2}t}{2}}   \hspace{2.0cm} \hbox{and} \hspace{2.0cm}  \tau=t
\label{xitau}
\end{equation}
where $a$, $b$ and $d$ are constants.  Without loss of generality we choose $a > 0$.
Hence:
$$
{\frac{\partial}{\partial r}} \rightarrow a{\frac{\partial}{\partial \xi}} \,\,\,\,\, , \hspace{1.0cm} {\frac{\partial^{3}}{\partial r^{3}}} \rightarrow a^{3}{\frac{\partial^{3}}{\partial \xi^{3}}}\,\,\,\,\, ,  \hspace{1.0cm} {\frac{\partial^{2}}{\partial z^{2}}} \rightarrow b^{2}{\frac{\partial^{2}}{\partial \xi^{2}}} \,\,\,\,\, ,
$$
\begin{equation}
{\frac{\partial^{2}}{\partial \varphi^{2}}} \rightarrow d^{2}{c_{s}}^{2}\varphi^{2}t^{2}{\frac{\partial^{2}}{\partial \xi^{2}}}-
d{c_{s}}t{\frac{\partial}{\partial \xi}} \,\,\,\,\, ,  \hspace{1.5cm}
{\frac{\partial}{\partial t}} \rightarrow {\frac{\partial}{\partial \tau}}-
d{\frac{{c_{s}}\varphi^{2}}{2}}{\frac{\partial}{\partial \xi}}
\label{xitauops}
\end{equation}
As a consequence we have:
\begin{equation}
\hat\rho_{1}(r,\varphi,z,t) \rightarrow \hat\rho_{1}(\xi,\tau)
\label{roumhattr}
\end{equation}
Using (\ref{xitauops}) and (\ref{roumhattr}) in (\ref{r3dckp}), since $a=d$, we find the
KdV equation in the $(\xi,\tau)$ space:
\begin{equation}
{\frac{\partial \hat\rho_{1}}{\partial \tau}}+\Bigg(ac_{s}+{\frac{b^{2}}{2a}}c_{s}\Bigg){\frac{\partial \hat\rho_{1}}{\partial \xi}}+a\alpha \hat\rho_{1} {\frac{\partial \hat\rho_{1}}{\partial \xi}}
+a^{3}\beta{\frac{\partial^{3} \hat\rho_{1}}{\partial \xi^{3}}}=0
\label{kdvxitau}
\end{equation}
which has the analytical soliton solution given by:
\begin{equation}
\hat\rho_{1}(\xi,\tau)={\frac{h_{1}}{h_{2}}}
sech^{2}\Bigg[{\frac{\sqrt{h_{1}}}{2}}\Bigg(\xi-u\tau\Bigg)\Bigg]
\label{kdvsol}
\end{equation}
where the constants are defined as:
\begin{equation}
h_{1}= {\frac{u-a{c_{s}}-b^{2}{c_{s}}/2a}{{a^{3}}\beta}} \hspace{2.0cm} \hbox{and} \hspace{2.0cm}
h_{2}= {\frac{\alpha}{3a^{2}\beta}}
\label{bees}
\end{equation}

The exact analytical soliton solution of (\ref{r3dckp}) in three cylindrical coordinates
is obtained substituting (\ref{xitau}) in (\ref{kdvsol}):
\begin{equation}
\hat\rho_{1}(r,\varphi,z,t)={\frac{h_{1}}{h_{2}}}
sech^{2}\Bigg\{{\frac{\sqrt{h_{1}}}{2}}\Bigg[ar+bz-\Bigg(u+a{\frac{{c_{s}}\varphi^{2}}{2}} \Bigg)t\Bigg]\Bigg\}
\label{ckpsol}
\end{equation}
where $u$ is a  parameter which satisfies $u>a{c_{s}}+b^{2}{c_{s}}/2a$  and the
phase velocity given by $u+a{\frac{{c_{s}}\varphi^{2}}{2}} $ is angle dependent.

The exact analytical soliton solution of the KP equation (\ref{kpxyztsual}) is  given by \cite{kp2004,kpsol,gino}:
\begin{equation}
\hat\rho_{1}(x,y,z,t)={\frac{3(U-w)}{{\mathcal{A}}\alpha}}
sech^{2}\Bigg[{\sqrt{{\frac{(U-w)}{4{\mathcal{A}}^{3}\beta}}}}\Bigg({\mathcal{A}}x+{\mathcal{B}}y+{\mathcal{C}}z-Ut\Bigg)\Bigg]
\label{kpsol}
\end{equation}
where ${\mathcal{A}}$, ${\mathcal{B}}$, ${\mathcal{C}}$ are real constants and  $w$ is given by:
\begin{equation}
w={\mathcal{A}}{c_{s}}+{\frac{{\mathcal{B}}^{2}{c_{s}}}{2}}+{\frac{{\mathcal{C}}^{2}{c_{s}}}{2}}
\label{pabove}
\end{equation}
We consider ${\mathcal{A}}>0$ and we have a  parameter $U$ such that $U > w$ .

\subsection{Conditions for the existence of  localized pulses}

\subsubsection{Cylindrical coordinates}

\begin{figure}[t]
\includegraphics[scale=0.7]{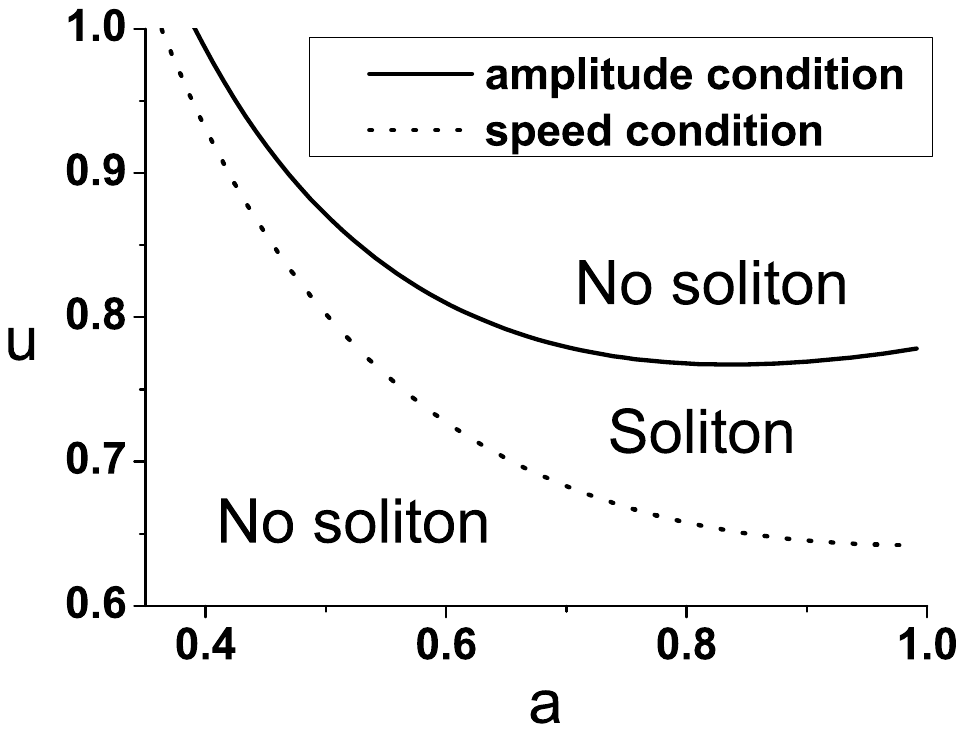}
\caption{Graphical representation of  (\ref{cond1}) (dashed line) and  (\ref{cond2}) (solid line).}
\label{fig1}
\end{figure}

The solution (\ref{ckpsol}) must be real and therefore  the constant $h_1$ must be positive. Moreover, following
Refs. \cite{kp2010,moslempp17} we assume that $a^2+b^2=1$ and hence:
\begin{equation}
u-a{c_{s}}- \frac{(1-a^2)c_{s}}{2a} > 0
\label{cond1}
\end{equation}
Since $\hat{\rho_1}$ is a normalized perturbation the following condition must hold:
\begin{equation}
\frac{h_1}{h_2} =  \frac{3}{a \alpha} \left(u - a c_s - \frac{(1-a^2) c_s}{2 a } \right)     < 1
\label{cond2}
\end{equation}
Within the region (in the $u - a$ plane)  where the conditions  (\ref{cond1}) and  (\ref{cond2})  are simultaneously
satisfied,  (\ref{ckpsol}) is well defined and we can have solitons.  This is illustrated in Fig. \ref{fig1}, where
we have chosen  $\rho_{0}= 1$ $fm^{-3}$ , $g = 1.15$ and  $m_G = 460$ MeV, which imply $c_s \simeq 0.64$. The stability
analysis can be made more rigorous with the introduction of the Sagdeev potential \cite{kp2010,moslempp17} which also
provides (\ref{cond1}) by using $\eta=\xi-ut=ar+bz-d{\frac{{c_{s}}\varphi^{2}t}{2}}-ut$ \ to rewrite equation (\ref{r3dckp})
as an energy balance equation.  For our present purposes the requirements  (\ref{cond1}) and  (\ref{cond2}) are sufficient.

\begin{figure}[ht!]
\begin{center}
\subfigure[ ]{\label{fig:first}
\includegraphics[width=0.48\textwidth]{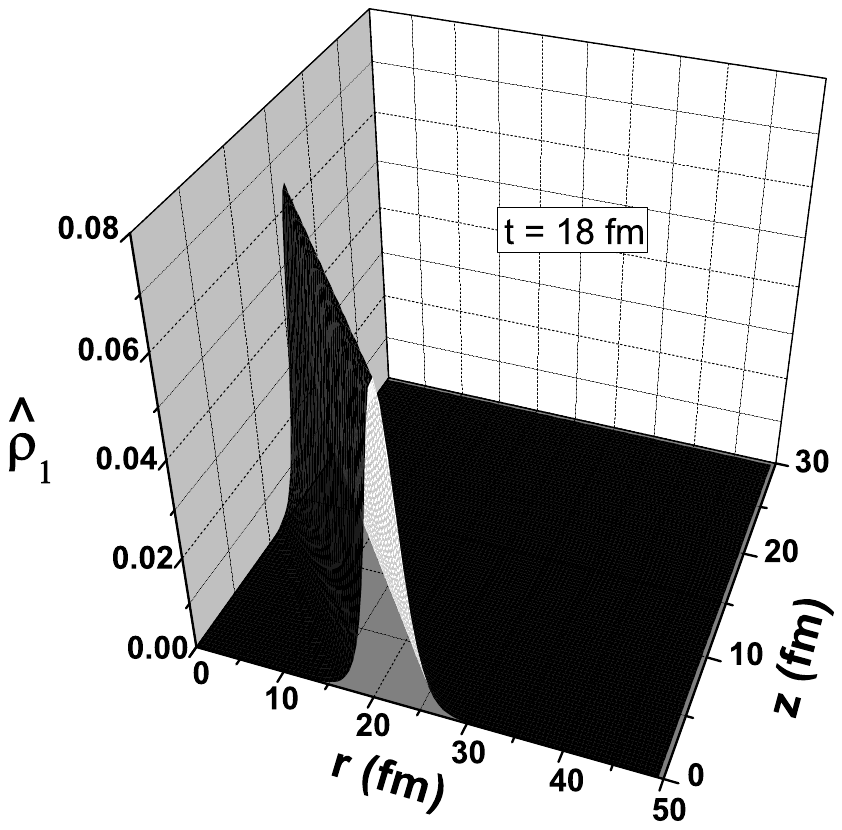}}
\subfigure[ ]{\label{fig:second}
\includegraphics[width=0.48\textwidth]{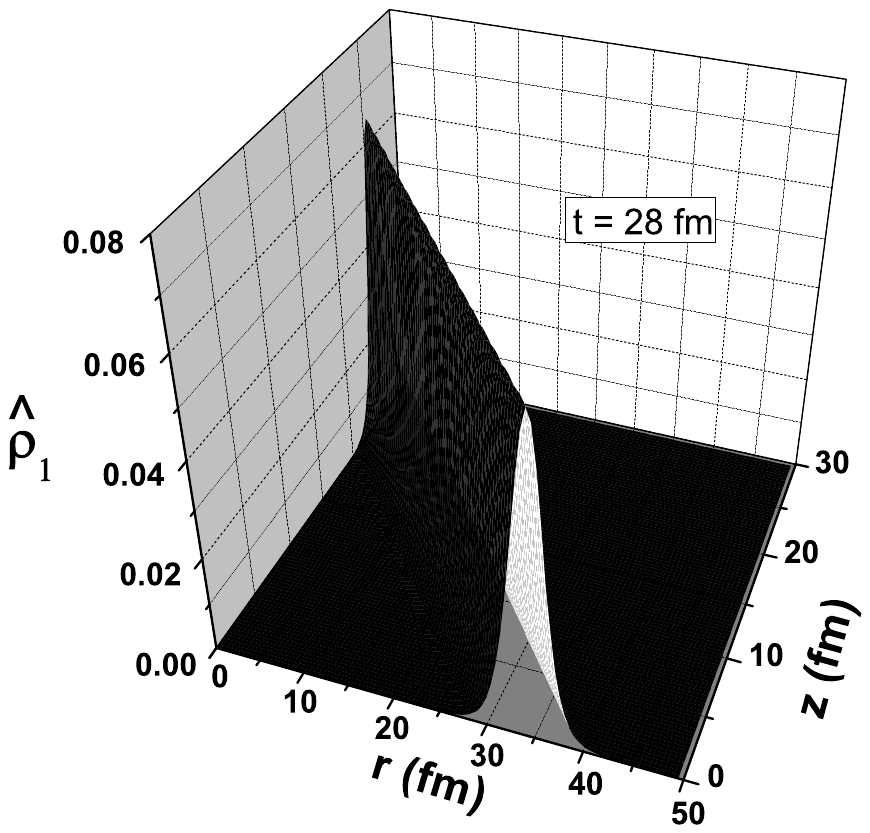}}\\
\subfigure[ ]{\label{fig:third}
\includegraphics[width=0.48\textwidth]{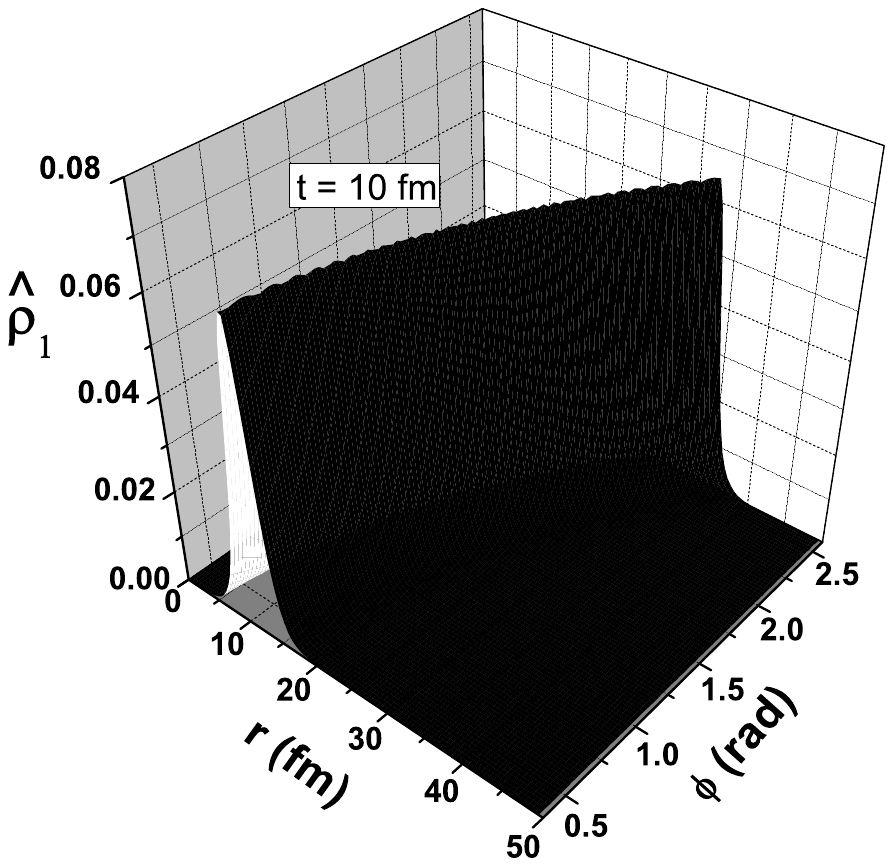}}
\subfigure[ ]{\label{fig:fourth}
\includegraphics[width=0.48\textwidth]{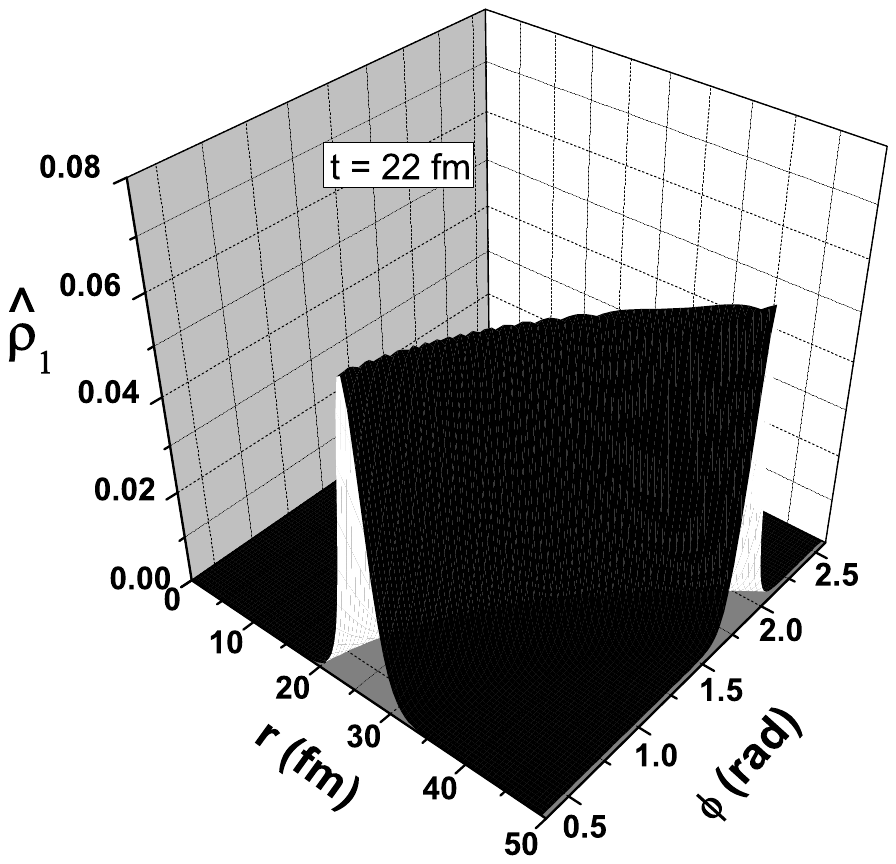}}
\end{center}
\caption{Graphical representation of  (\ref{ckpsol}) for different times, increasing from the left to the right. Upper and
lower plots are for different  parameter choices (see text). }
\label{fig2}
\end{figure}

An example of soliton evolution is presented in  Fig. \ref{fig2}.
We show a plot of (\ref{ckpsol}) with fixed $\varphi=0^{o}$,  $a=0.6$, $b=0.8$, $u=0.73$ and $z$ varying in the range
$0 \, \mbox{fm}  \leq z \,   \leq  30 \,  \mbox{fm} $. This choice of parameters  satisfies  the soliton conditions  (\ref{cond1}) and
(\ref{cond2}).  The pulse is observed at two times:  $t=18$ fm in Fig. \ref{fig:first}  and at $t=28$ fm in Fig. \ref{fig:second}.
From the figure we can see that the cylindrical pulse expands outwards in the radial direction. The regions with larger $z$
expand with a delay with respect to the central ($z=0$) region.

Keeping $z=1$ fm fixed,  we show the time evolution of  (\ref{ckpsol}) from $t=10$ fm (Fig. \ref{fig:third}) to
$t=22$ fm  (Fig.  \ref{fig:fourth}).  The azimuthal angle varies in the range  $20^{o}$ $\leq \varphi \leq $ $150^{o}$.
From the parenthesis in   (\ref{ckpsol}) we can see that the expansion velocity grows with the angle. This asymmetry can be
clearly seen in the figure, where the large angle ``backward'' region moves faster the small angle ``forward'' region. The breaking
of $z$ invariance and  azimuthal symmetry is entangled with the soliton stability and with the physical properties of the system
(contained in the parameters $h_1$, $h_2$ and $c_s$).

\subsubsection{Cartesian coordinates}

We perform the study of the existence condition for the solution (\ref{kpsol}), which must be real and therefore  the constant $U-w$ must be positive.  Again we have chosen
$\rho_{0}= 1$ $fm^{-3}$ , $g = 1.15$ and  $m_G = 460$ MeV, which imply $c_s \simeq 0.64$.
We also set ${\mathcal{C}}=0.5$ and  extend the condition in
Refs. \cite{kp2010,moslempp17} to ${\mathcal{A}}^2+{\mathcal{B}}^2+{\mathcal{C}}^{2}=1$ .  As mentioned, $U > w$ and from (\ref{pabove}):
\begin{equation}
U-{\mathcal{A}}{c_{s}}-
{\frac{[1-{\mathcal{A}}^{2}-{(0.5)}^{2}]{c_{s}}}{2}}-{\frac{{(0.5)}^{2}{c_{s}}}{2}}
> 0
\label{cond3}
\end{equation}

\begin{figure}[t]
\includegraphics[scale=0.7]{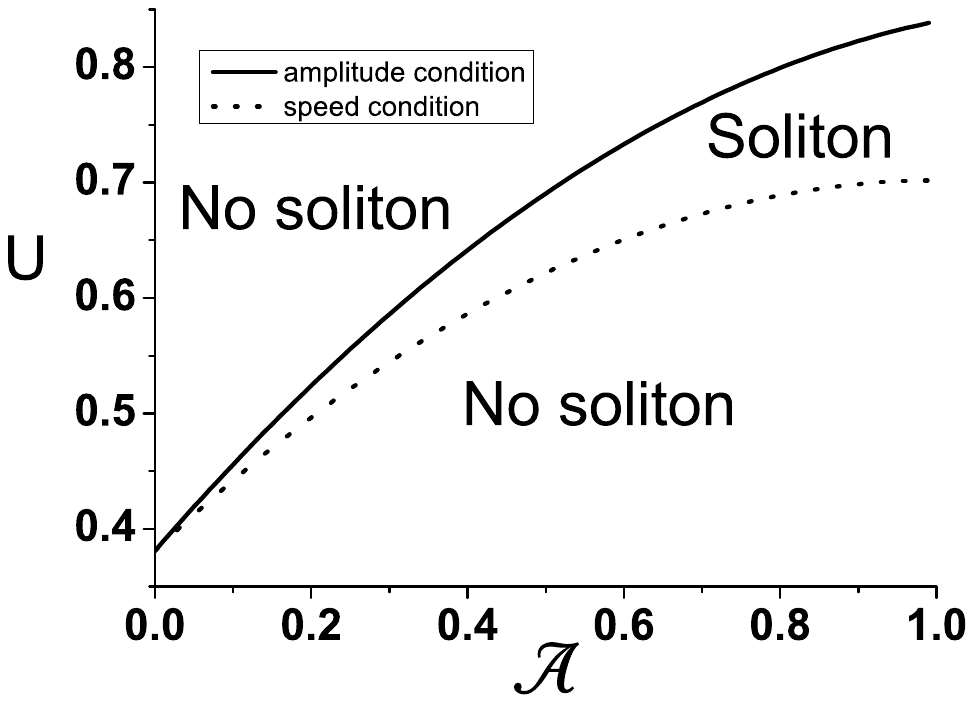}
\caption{Graphical representation of  (\ref{cond3}) (dashed line) and  (\ref{cond4}) (solid line).}
\label{fig3}
\end{figure}

Again, $\hat{\rho_1}$ is a normalized perturbation, so the amplitude condition must hold:
\begin{equation}
{\frac{3(U-w)}{{\mathcal{A}}\alpha}} =  \frac{3}{{\mathcal{A}} \alpha} \left(U-{\mathcal{A}}{c_{s}}-
{\frac{[1-{\mathcal{A}}^{2}-{(0.5)}^{2}]{c_{s}}}{2}}-{\frac{{(0.5)}^{2}{c_{s}}}{2}}\right)     < 1
\label{cond4}
\end{equation}
Within the region (in the $U - {\mathcal{A}}$ plane)  where the conditions  (\ref{cond3}) and  (\ref{cond4})  are simultaneously
satisfied,  (\ref{kpsol}) is well defined and we can have solitons as it can be seen in Fig. \ref{fig3}.  Again, the stability analysis can be performed more rigorously with the introduction of the Sagdeev potential \cite{kp2010,moslempp17}, which also
provides (\ref{cond3}) by using ${\mathcal{A}}x+{\mathcal{B}}y+{\mathcal{C}}z-Ut$, to rewrite equation (\ref{kpxyztsual}) as an energy balance equation.  The requirements (\ref{cond3}) and  (\ref{cond4}) are sufficient
to provide a soliton propagation in the present case.

\begin{figure}[ht!]
\begin{center}
\subfigure[ ]{\label{fig:firsto}
\includegraphics[width=0.48\textwidth]{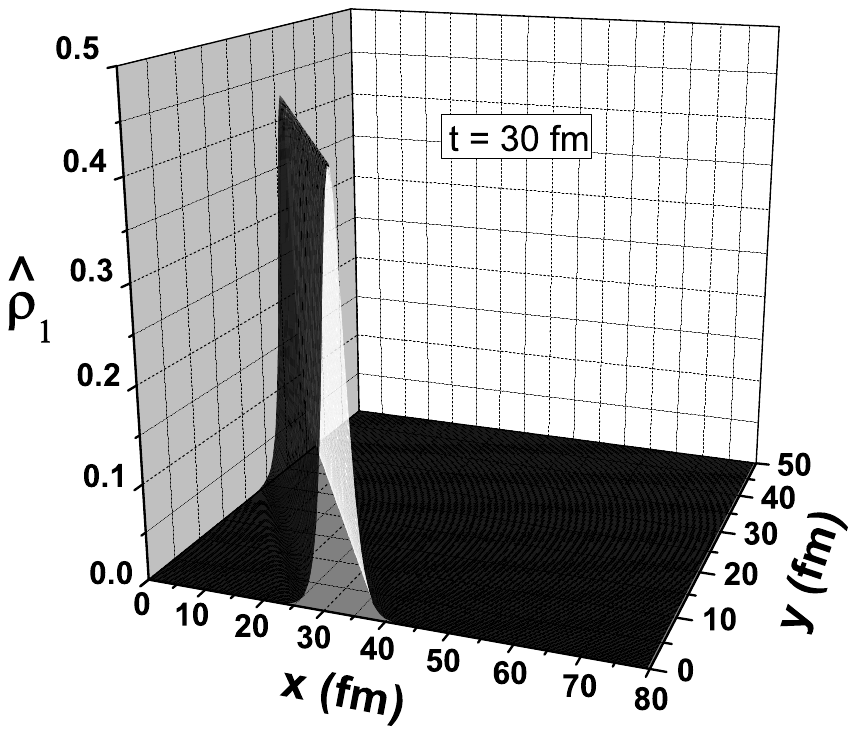}}
\subfigure[ ]{\label{fig:secondo}
\includegraphics[width=0.48\textwidth]{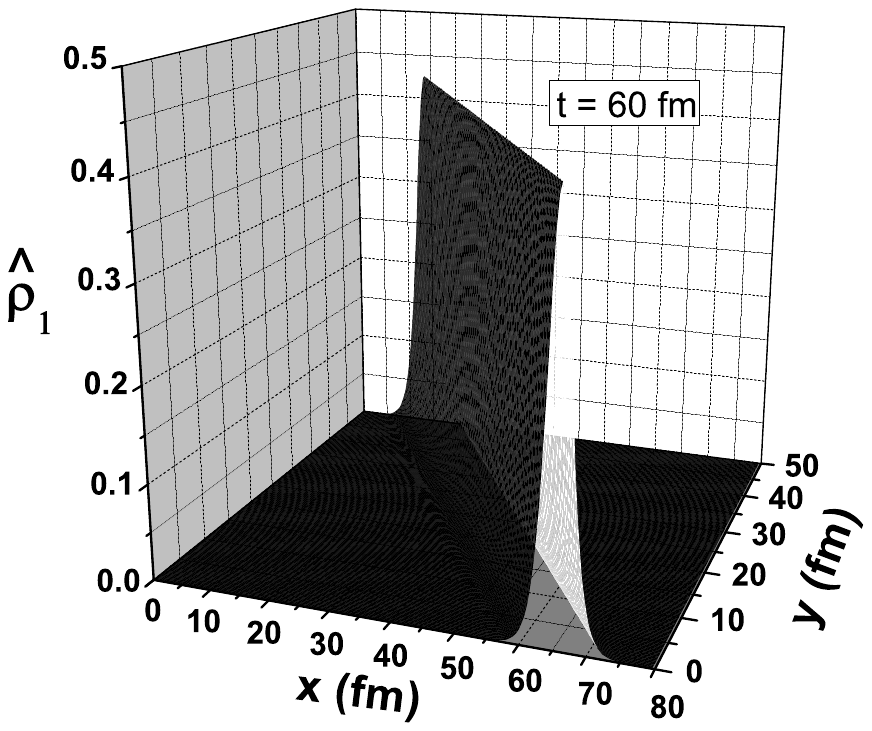}}\\
\subfigure[ ]{\label{fig:thirdo}
\includegraphics[width=0.48\textwidth]{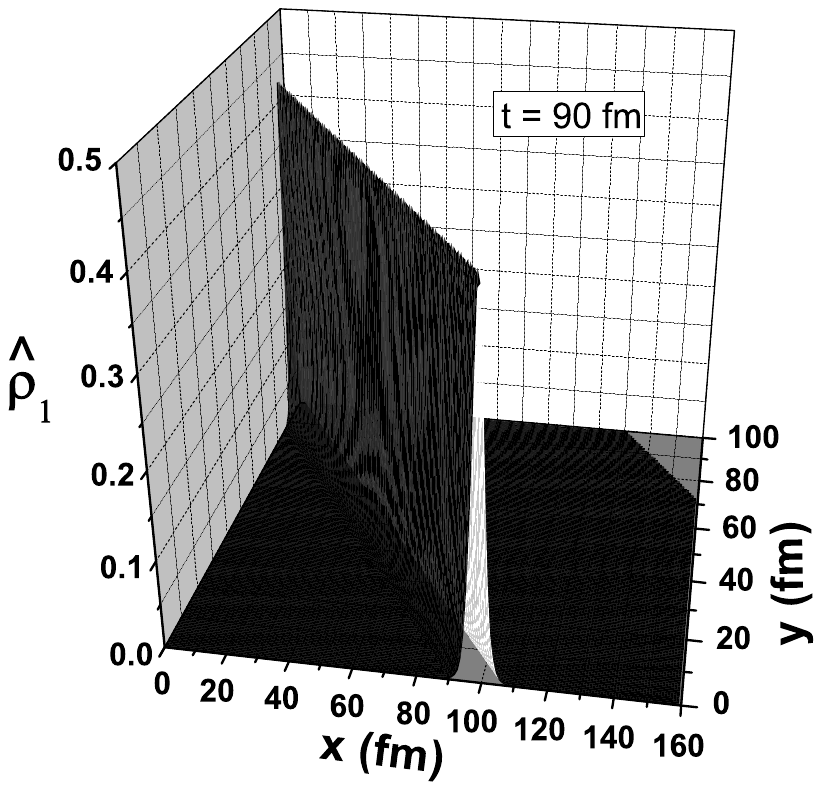}}
\subfigure[ ]{\label{fig:fourtho}
\includegraphics[width=0.48\textwidth]{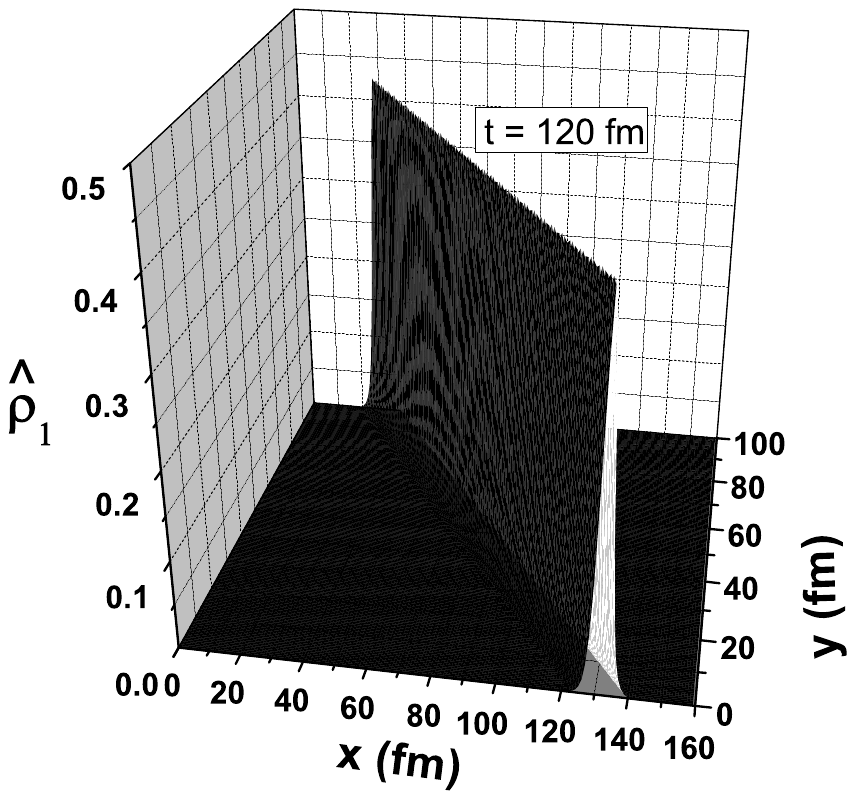}}
\end{center}
\caption{Graphical representation of (\ref{kpsol}) for different times, increasing from the left to the right and from upper to the
lower. The plots are for the same parameter choices (see text). }
\label{fig4}
\end{figure}

A simple example of soliton evolution is presented in  Fig. \ref{fig4}.
We show a plot of (\ref{kpsol}) with fixed $z=1$ fm,  ${\mathcal{A}}=0.6$, ${\mathcal{B}} \cong 0.62$, $U=0.66$ and $y$ varying in the range
$0 \, \mbox{fm}  \leq y \,   \leq  50 \,  \mbox{fm} $. This choice of parameters  satisfies  the soliton conditions  (\ref{cond3}) and
(\ref{cond4}).  The pulse is observed at four times:  $t=30$ fm (Fig. \ref{fig:firsto})  to $t=120$ fm (Fig. \ref{fig:fourtho}).
From the figure we can see that the cartesian pulse expands outwards in the $x$ direction
keeping its shape and form.

\section{Conclusions}

We have described in detail how to obtain a KP equation in three dimensions in cylindrical
and cartesian coordinates in the context of relativistic fluid dynamics of a cold quark gluon plasma.  To this end, we have
used the equation of state derived from QCD in \cite{nos2011}.  The resulting nonlinear relativistic wave equations are for small
perturbations in the baryon density.

For the cartesian KP the exact soliton solution is a supersonic bump keeping its shape without deformation. The cartesian KP contains some particular cases such as KdV and the breaking wave equation already encountered in our previous works
\cite{nos2010,nos2011a}.  For the cylindrical KP (cKP) we also have an exact supersonic soliton solution which deforms slightly as time goes on due to the angular dependence in the phase.

We conclude that relativistic fluid dynamics supports nonlinear solitary waves even with the inclusion of transverse perturbations in cylindrical and cartesian geometry.

\section{Appendix}

In this appendix we start from the equations of relativistic hydrodynamics and, using the linearization approximation, we derive a wave equation for
perturbations in the pressure. This equation has travelling wave solutions which represent acoustic waves. In the derivation presented here we follow
closely Ref. \cite{hidro1}. The energy density and pressure for the relativistic fluid are written as:
\begin{equation}
\varepsilon(\vec{r},t)=\varepsilon_{0}+\delta \varepsilon(\vec{r},t)
\label{15}
\end{equation}
and
\begin{equation}
p(\vec{r},t)=p_{0}+\delta p(\vec{r},t)
\label{16}
\end{equation}
respectively.  The uniform relativistic fluid is defined by $\varepsilon_{0}$ and $p_{0}$,  while
$\delta \varepsilon$ and $\delta p$ correspond to perturbations in this fluid.  Energy-momentum conservation implies that:
\begin{equation}
\partial_{\mu}T^{\mu\nu}=0
\label{7}
\end{equation}
where $T^{\mu\nu}$ is the energy-momentum tensor  given by:
\begin{equation}
T^{\mu\nu}=(\varepsilon+p)u^{\mu}u^{\nu}-pg^{\mu\nu}
\label{enermomtensor}
\end{equation}
Linearization consists in keeping only first order terms such as $\delta \varepsilon$, $\delta P$ and $\vec{v}$ and
neglect terms proportional to:
\begin{equation}
{{v}}^{2}, \hspace{0.2cm} v\delta \varepsilon, \hspace{0.2cm} v \delta P, \hspace{0.2cm}
\vec{v} \cdot \vec{\nabla}v, \hspace{0.2cm} (\vec{v} \cdot \vec{\nabla}) \vec{v}
\label{aplin}
\end{equation}
and also neglect higher powers of these products or other combinations  of  them.
Naturally we have:
\begin{equation}
\gamma={\frac{1}{\sqrt{1-v^{2}}}} \sim 1
\label{aplina}
\end{equation}
From (\ref{7}) we have:
\begin{equation}
u^{\mu}\partial_{\nu}[(\varepsilon + p)u^{\nu}]+(\varepsilon + p)u^{\nu}\partial_{\nu}u^{\mu}-\partial_{\nu}(pg^{\nu\mu})=0
\label{8ag}
\end{equation}
The temporal component ($\mu=0$) of the above equation is given by:
\begin{equation}
\gamma\partial_{0}[(\varepsilon + p)\gamma]+\gamma\partial_{i}[(\varepsilon + p)u^{i}]+
(\varepsilon + p)u^{0}\partial_{0}\gamma+(\varepsilon + p)u^{i}\partial_{i}\gamma-\partial_{0}p=0
\label{17}
\end{equation}
which, after using (\ref{aplin}) and (\ref{aplina}),   becomes:
$$
\partial_{0}(\varepsilon + p)+\partial_{i}[(\varepsilon + p)v^{i}]-\partial_{0}p=0
$$
or
\begin{equation}
{\frac{\partial \varepsilon}{\partial t}}+ \vec{\nabla} \cdot [(\varepsilon + p)\vec{v}]=0
\label{energcons}
\end{equation}
For the $j$-th spatial component ($\mu=j$)  in (\ref{8ag}) we have:
$$
u^{j}\partial_{0}[(\varepsilon + p)u^{0}]
+u^{j}\partial_{i}[(\varepsilon + p)u^{i}]+(\varepsilon + p)u^{0}\partial_{0}u^{j}+
(\varepsilon + p)u^{i}\partial_{i}u^{j}-\partial^{j}p=0
$$
which, with the  use of (\ref{aplin}),  becomes:
\begin{equation}
{\frac{\partial}{\partial t}}[(\varepsilon + p)\vec{v}]+ \vec{\nabla}p=0
\label{newtonl}
\end{equation}
Substituting the expansions (\ref{15}) and (\ref{16}) in (\ref{energcons}) and (\ref{newtonl}) we find:
\begin{equation}
{\frac{\partial }{\partial t}}[\varepsilon_{0}+\delta \varepsilon]+
\vec{\nabla} \cdot [(\varepsilon_{0}+\delta \varepsilon + p_{0}+\delta p) \vec{v}]=0
\label{energconsexp}
\end{equation}
and
\begin{equation}
{\frac{\partial}{\partial t}}[(\varepsilon_{0}+\delta \varepsilon + p_{0}+\delta p)\vec{v}]+
\vec{\nabla}[p_{0}+\delta p]=0
\label{newtonlexp}
\end{equation}
Using the linearization (\ref{aplin}) and (\ref{aplina}) in (\ref{energconsexp}) and (\ref{newtonlexp}) they become:
\begin{equation}
{\frac{\partial (\delta \varepsilon)}{\partial t}}+
(\varepsilon_{0}+p_{0})\vec{\nabla} \cdot  \vec{v}=0
\label{enerlin}
\end{equation}
and
\begin{equation}
(\varepsilon_{0}+p_{0}){\frac{\partial \vec{v}}{\partial t}}+
\vec{\nabla}(\delta p)=0
\label{newtolin}
\end{equation}
Equation (\ref{enerlin}) expresses  energy conservation and equation (\ref{newtolin}) is  Newton's second law.
Integrating (\ref{newtolin}) with respect to the time and setting the integration constant to zero we find:
\begin{equation}
\vec{v}=-{\frac{1}{(\varepsilon_{0}+p_{0})}}\int{\vec{\nabla}(\delta p)} dt
\label{velolin}
\end{equation}
which inserted in (\ref{enerlin}) yields:
\begin{equation}
{\frac{\partial (\delta \varepsilon)}{\partial t}}-
\int \vec{\nabla}^{2}(\delta p)dt=0
\label{enerlina}
\end{equation}
Performing the time derivative we obtain:
\begin{equation}
{\frac{\partial^{2} (\delta \varepsilon)}{\partial t^{2}}}-
\vec{\nabla}^{2}(\delta p)=0
\label{enerlinaa}
\end{equation}
Assuming that
\begin{equation}
\delta \varepsilon={\frac{\partial \varepsilon}{\partial p}} \delta p
\label{presom}
\end{equation}
with $\partial \varepsilon / \partial p$ being a constant, we have (\ref{enerlinaa}) rewritten  as:
\begin{equation}
{\frac{\partial \varepsilon}{\partial p}}{\frac{\partial^{2} (\delta p)}{\partial t^{2}}}
-\vec{\nabla}^{2}(\delta p)=0
\label{enerlinaaa}
\end{equation}
The above equation is a wave equation from where we can identify the velocity of propagation as:
\begin{equation}
c_{s}=\bigg({\frac{\partial p}{\partial \varepsilon}}\bigg)^{1/2}
\label{som}
\end{equation}
where $c_{s}$ is the speed of sound. Equation  (\ref{enerlinaaa}) can then be finally  written as:
\begin{equation}
\vec{\nabla}^{2}(\delta p)
-{\frac{1}{{c_{s}}^{2}}}{\frac{\partial^{2} (\delta p)}{\partial t^{2}}}=0
\label{enerlinfinal}
\end{equation}
which describes the  propagation of a pressure wave  in the fluid.

The derivation presented above shows that the existence of sound waves in a relativistic
perfect fluid depends only on the equation of state $p=p(\varepsilon)$.  In particular,
these formulas show that we can have acoustic waves in a medium made of massless particles.
As a simple example, let us consider the equation of state given by (\ref{eps}) and (\ref{pres})
in the case where we have  no gluons ($g=0$) and only massless quarks. In this case (\ref{eps}) and
(\ref{pres})  reduce to:
\begin{equation}
\varepsilon=\mathcal{B}_{QCD}
+3{\frac{\gamma_{Q}}{2{\pi}^{2}}}{\frac{{k_{F}}^{4}}{4}}
\label{epsng}
\end{equation}
and
\begin{equation}
p=-\mathcal{B}_{QCD}
+{\frac{\gamma_{Q}}{2{\pi}^{2}}}{\frac{{k_{F}}^{4}}{4}}
\label{presng}
\end{equation}
which can be combined to give:
\begin{equation}
p={\frac{1}{3}}\varepsilon-{\frac{4}{3}}\mathcal{B}_{QCD}
\label{eosqg}
\end{equation}
with the speed of sound  $c_{s}$  given by  (\ref{som}):
\begin{equation}
{c_{s}}^{2}={\frac{\partial p}{\partial \varepsilon}}={\frac{1}{3}}
\label{soundone}
\end{equation}

\begin{acknowledgments}
We are deeply grateful to R. A. Kraenkel for useful discussions.
This work was  partially financed by the Brazilian funding
agencies CAPES, CNPq and FAPESP.
\end{acknowledgments}

\end{document}